\documentclass[twocolumn,showpacs,preprintnumbers,amsmath,amssymb]{revtex4}
\usepackage{graphicx}
\usepackage{dcolumn}
\usepackage{bm}

\begin{document}
\title{Transient Zitterbewegung of charge carriers in graphene and carbon nanotubes}
\date{\today}

\author{Tomasz M. Rusin}
\email{Tomasz.Rusin@centertel.pl}
\author{Wlodek Zawadzki\dag}
\affiliation{*PTK Centertel Sp. z o.o., ul. Skierniewicka 10A, 01-230 Warsaw, Poland\\
         \dag Institute of Physics, Polish Academy of Sciences, Al. Lotnik\'ow 32/46, 02-688 Warsaw, Poland}


\begin{abstract}
Observable effects due to trembling motion (Zitterbewegung, ZB) of charge carriers in bilayer
graphene, monolayer graphene and carbon nanotubes are calculated. It is shown that,
when the charge carriers are prepared in the form of gaussian wave packets, the ZB has a transient
character with the decay time of femtoseconds in graphene and picoseconds in nanotubes.
Analytical results for bilayer graphene allow us to investigate phenomena which
accompany the trembling motion. In particular, it is shown that the transient character of ZB
in graphene is due to the fact that wave subpackets related to positive and negative
electron energies move in opposite directions, so their overlap diminishes with time.
This behavior is analogous to that of the wave packets representing relativistic
electrons in a vacuum.
\end{abstract}

\maketitle
\section{Introduction}
The trembling motion (Zitterbewegung, ZB), first devised
by Schroedinger for free relativistic electrons in a vacuum
\cite{Schroedinger30}, has become in the last two years subject of
great  theoretical interest as it has turned out that this
phenomenon should occur in many situations in semiconductors
\cite{Cannata90,Zawadzki05KP,Zawadzki06,Schliemann05,Katsnelson06,Rusin07,
Cserti06,Winkler06,Trauzettel07}. Whenever one deals with two or more energy branches,
an interference of the corresponding upper and lower energy states
results in the trembling motion even in the absence of external
fields. Due to a formal similarity between two interacting bands in a solid and the Dirac
equation  for relativistic electron in a vacuum one can use methods developed in the relativistic
quantum mechanics for non-relativistic electrons in solids \cite{ZawadzkiHMF,ZawadzkiOPS}.
Most of the ZB studies for semiconductors
took as a starting point plane electron waves
(see, however, Refs. \cite{Huang52,Lock79,Schliemann05,Rusin07}).
On the other hand, Lock \cite{Lock79} in his important paper observed:
'Such a wave is not localized and it seems to be of a limited practicality to speak of
rapid fluctuations in the average position of a wave of infinite extent.'
Using the Dirac equation Lock showed that, when an electron is represented by a wave packet,
the ZB oscillations do not remain undamped but become
transient. In particular, the disappearance of oscillations at sufficiently large times is
guaranteed by the Riemann-Lebesgue theorem as long as the wave packet is a smoothly varying function.
Since the ZB is by its nature not a stationary state but a dynamical phenomenon,
it is natural to study it with the use of wave packets. These have become a practical instrument when
femtosecond pulse technology emerged (see Ref. \cite{Garraway95}).

In the following we study theoretically the Zitterbewegung of mobile charge carriers in three
modern materials: bilayer graphene, monolayer graphene and carbon nanotubes. We have three objectives
in mind. First, we obtain for the first time analytical results for the ZB of gaussian wave packets
which allows us to study not only the trembling motion itself but also effects that accompany
this phenomenon. Second, we describe for the first time the transient character of ZB
in solids, testing on specific examples the general predictions of Ref. \cite{Lock79}. Third, we
look for observable phenomena and select both systems and parameters which appear most
promising for experiments.
We first present our analytical results for bilayer graphene and then quote some predictions
for observable quantities in monolayer graphene  and carbon nanotubes.

\section{Bilayer graphene}
Two-dimensional Hamiltonian for bilayer graphene is well approximated by \cite{McCann06PRL}
\begin{equation} \label{BG_H}
 \hat{H}_{B} = -\frac{1}{2m^*}\left(\begin{array}{cc}
     0 & (p_x-ip_y)^2 \\  (p_x+ip_y)^2 & 0 \\     \end{array}\right),
\end{equation}
where $m^*=0.054m_0$. The form (\ref{BG_H}) is valid for energies
$2\ {\rm meV}<{\cal E}<100\ $meV in the conduction band.
The energy spectrum is ${\cal E}=\pm E$, where $E=\hbar^2k^2/2m^*$, i.e.
there is no energy gap between the conduction and valence bands. The position operator in the
Heisenberg picture is a $2\times 2$ matrix
$\hat{x}(t) = \exp(i\hat{H}_Bt/\hbar)\hat{x}\exp(-i\hat{H}_Bt/\hbar).$
We calculate
\begin{equation} \label{BG_x_11}
\hat{x}_{11}(t)= \hat{x}  + \frac{k_y}{k^2}\left[ 1 -\cos\left(\frac{\hbar k^2t}{m^*}\right)\right],
\end{equation}
where $k^2=k_x^2+k_y^2$.
The third term represents the Zitterbewegung with the frequency
$\hbar\omega_Z= 2\hbar^2k^2/2m^*$, corresponding to the energy difference
between the upper and lower energy branches for a given value of $k$.

We want to calculate the ZB of a charge carrier represented by a two-dimensional wave packet
\begin{equation} \label{BG_psi}
 \psi(\bm {r},0)= \frac{1}{2\pi}\frac{d}{\sqrt{\pi}}
    \int d^2 \bm k e^{-\frac{1}{2}d^2k_x^2-\frac{1}{2}d^2(k_y-k_{0y})^2} e^{i\bm k \bm r}
 \left(\begin{array}{c} 1 \\ 0 \end{array}\right).
\end{equation}
The packet is centered at $\bm k_0 = (0,k_{0y})$ and is characterized by a width $d$.
The unit vector $(1,0)$ is a convenient choice, selecting
the [11] component of $\hat{x}(t)$, see Eq. (\ref{BG_x_11}). An average of $\hat{x}_{11}(t)$ is a
two-dimensional integral which we calculate analytically
\begin{equation} \label{BG_x(t)}
\bar{x}(t) = \langle \psi(\bm {r})|\hat{x}(t)| \psi(\bm {r})\rangle
           = \bar{x}_c + \bar{x}_Z(t)
\end{equation}
where $\bar{x}_c = (1/k_{0y}) \left[1 - \exp(-d^2k_{0y}^2)\right]$, and
\begin{eqnarray} \label{BG_xZ(t)}
\bar{x}_Z(t) = \frac{1}{k_{0y}} \left[  \exp\left(-\frac{\delta^4 d^2 k_{0y}^2}{d^4+\delta^4}\right)
               \cos\left(\frac{\delta^2d^4k_{0y}^2}{d^4+\delta^4}  \right) \right.  \nonumber \\
             \left. -\exp(-d^2k_{0y}^2) \right], \ \ \ \ \
\end{eqnarray}
in which $\delta = \sqrt{\hbar t/m^*}$ contains the time dependence. We enumerate the main features of ZB
following from Eqs. (\ref{BG_x(t)}) and (\ref{BG_xZ(t)}). First, in order to have
the ZB in the direction $x$ one needs an initial transverse momentum $\hbar k_{0y}$.
Second, the ZB frequency depends only weakly on the packet width:
$\omega_Z=(\hbar k_{0y}^2/m^*)(d^4/(d^4+\delta^4)$, while its amplitude is strongly dependent
on the width $d$.
Third, the ZB has a transient character,
as it is attenuated by the exponential term. For small $t$ the amplitude of $\bar{x}_Z(t)$ diminishes
as $\exp(-\Gamma_Z^2t^2)$ with
\begin{equation} \label{BG_GammaZ}
\Gamma_Z= \frac{\hbar k_{0y}}{m^*d}.
\end{equation}
Fourth, as $t$ (or $\delta$) increases
the cosine term tends to unity and the first term in Eq. (\ref{BG_xZ(t)}) cancels out with the second term,
which illustrates the Riemann-Lebesgue theorem (see Ref. \cite{Lock79}).
After the oscillations disappear, the charge carrier is displaced by the amount $\bar{x}_c$, which is a
'remnant' of ZB. Fifth, for very narrow packets ($d\rightarrow\infty$) the exponent in Eq. (\ref{BG_xZ(t)})
tends to unity, the oscillatory term is $\cos(\delta^2k_{0y}^2)$ and the last term vanishes.
In this limit we recover undamped ZB oscillations.

\begin{figure}
\includegraphics[width=8.5cm,height=11.33cm]{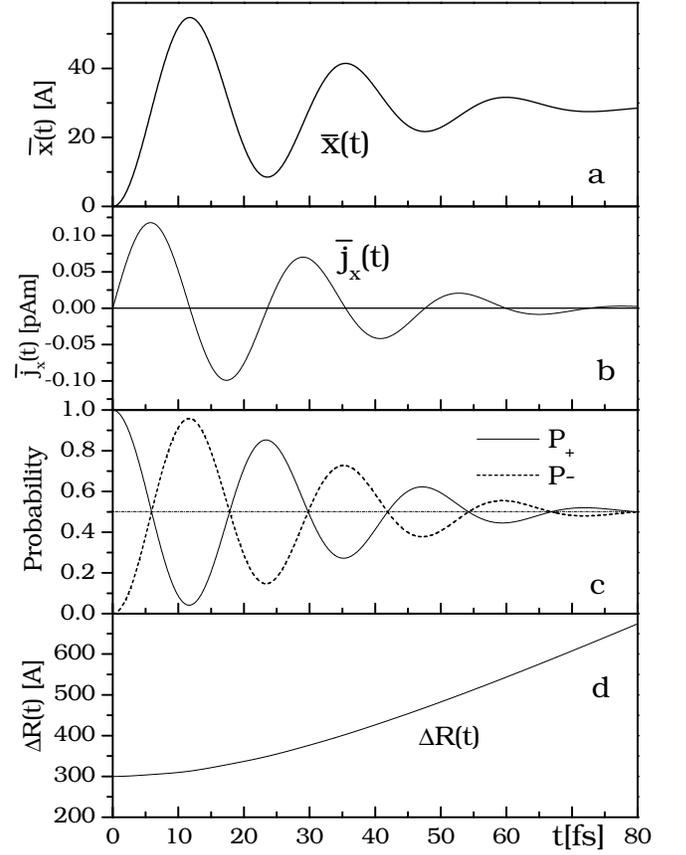}
\caption{ \label{Fig1} Zitterbewegung of a charge carrier in bilayer graphene {\it versus}
time, calculated for a gaussian wave packet width $d=300$\AA\ and $k_{0y}=3.5\times 10^8$m$ ^{-1}$:
a) displacement, b) electric current,
c) probability densities for upper and lower components of the wave function,
d) dispersion $\Delta R(t)$.
The decay time is $\Gamma_Z^{-1}=40$ fs, see Eq. (\ref{BG_GammaZ}).}
\end{figure}

Next, we consider observable quantities related to the ZB, beginning by the current.
The latter is given by the velocity multiplied by charge.
The velocity is simply $\bar{v}_x=\partial \bar{x}_Z/\partial t$,
where $\bar{x}_Z$ is given by Eq. (\ref{BG_xZ(t)}).
The calculated current is plotted in Fig. \ref{Fig1}b,
its oscillations are a direct manifestation of ZB.
The Zitterbewegung is also accompanied by a time dependence of upper and lower components of the wave
function. To characterize this evolution we define probability densities for the upper and lower
components
\begin{equation} \label{BG_def_Ppm}
P_{\pm}(t) = \langle \psi(\bm r, t) \left|\frac{1\pm \hat{\beta}}{2}\right| \psi(\bm r,t) \rangle,
\end{equation}
where $\hat{\beta} =\left(\begin{array}{cc} 1 & 0 \\  0 & -1 \\ \end{array}\right)$,
and the time-dependent wave function is $\psi(\bm r,t)=\exp(-i\hat{H}_Bt/\hbar)\psi(\bm r,0)$.
We have
\begin{eqnarray} \label{BG_psit}
 \psi(\bm {r},t)= \frac{1}{2\pi}\frac{d}{\sqrt{\pi}}
    \int d^2 \bm k e^{-\frac{1}{2}d^2k_x^2-\frac{1}{2}d^2(k_y-k_{0y})^2} e^{i\bm k \bm r}
    \times \nonumber \\
 \left(\begin{array}{c} \cos(\hbar k^2t/2m) \\
   i(k_+/k)^2\sin(\hbar k^2t/2m) \end{array}\right), \ \
\end{eqnarray}
where $k_+=k_x+ik_y$. For $t=0$ Eq. (\ref{BG_psit}) reduces to Eq. (\ref{BG_psi}).
The calculated probability densities are
\begin{eqnarray} \label{BG_Ppm}
P_{\pm}(t) = \frac{1}{2} \pm \frac{1}{2}\frac{d^2}{s^4}
    \exp\left(-\frac{\delta^4d^2k_{0y}^2}{s^4} \right)  \nonumber \\
     \times \left[d^2      \cos\left(\frac{\delta^2d^4k_{0y}^2}{s^4} \right)  -
            \delta^2 \sin\left(\frac{\delta^2d^4k_{0y}^2}{s^4} \right)\right],
\end{eqnarray}
where $s^4=d^4+\delta^4$.
The time dependence of $P_{\pm}(t)$ is illustrated in Fig. \ref{Fig1}c.
Clearly, there must be
$P_+(t)+ P_-(t)=1$ at any time, but it is seen that the probability density 'flows' back
and forth between the two components. It is clear that the oscillating probability is
directly related to ZB. Since in bilayer graphene the upper and lower components are associated
with the first or second layer, respectively \cite{McCann06NP}, it follows that, at least for this system,
the trembling motion represents oscillations of a charge carrier between the two graphene layers.
For sufficiently long times there is $P_{\pm}=1/2$, so that the final probability is equally distributed
between the two layers. For a very narrow packet ($d\rightarrow \infty$) we have
$P_{\pm}=(1/2)[1\pm \cos^2(\delta^2k_{0y}^2)]$, which indicates that the probability oscillates
without attenuation. For $k_{0y}=0$ there is $P_{\pm}=(1/2)[1\pm d^4/s^4]$, i.e.
there are no oscillations and the initial probability $(1,0)$ simply decays into $(1/2,1/2)$.

The above phenomenon can be considered from the point of view of the entropy:
$S=-P_+\lg_2P_+$$-P_-\lg_2P_-$.
At the beginning (the carrier is in one layer) the entropy is zero, at the end (when the probability
is equally distributed in two layers) the entropy is $\lg_22$. However, the entropy increases
in the oscillatory fashion (see Ref. \cite{Novaes04}).

The transient character of ZB is accompanied by a temporal spreading of the wave packet.
In fact, the question arises whether the damping of ZB is not simply {\it caused} by the
spreading of the packet. To study this question we calculate an explicit form of the wave function
given by integrals (\ref{BG_psit}). The result is
\begin{eqnarray} \label{BG_psi_up}
\psi^{up}(\bm r, t) = \frac{d}{\sqrt{\pi}s^4}
\exp\left(-\frac{d^2\rho^2}{2s^4}\right)
\exp\left(-\frac{\delta^4d^2k_{0y}^2}{2s^4}\right)   \nonumber \\
\times
\exp\left(\frac{iyd^4k_{0y}}{s^4}\right)
\left[     d^2\cos\left(\frac{\delta^2\rho_{k0}^2}{2s^4}\right) +
   \delta^2\sin\left(\frac{\delta^2\rho_{k0}^2}{2s^4}\right) \right], \ \ \
\end{eqnarray}
\begin{eqnarray} \label{BG_psi_low}
\psi^{low}(\bm r, t) = \frac{-id}{\sqrt{\pi}s^4} \exp\left(-\frac{d^2\rho^2}{2s^4}\right)
\exp\left(-\frac{\delta^4d^2k_{0y}^2}{2s^4}\right) \nonumber \\
\times
\exp\left(\frac{iyd^4k_{0y}}{s^4}\right) \left( \frac{x+iy+d^2k_{0y}}{\rho_{k0}} \right)^2 \nonumber
\end{eqnarray}
\begin{eqnarray}
\times
\left[\left(\frac{2s^4}{\rho_{k0}^2} + d^2\right)
  \sin\left(\frac{\delta^2\rho_{k0}^2}{2s^4}\right)
 -\delta^2\cos\left(\frac{\delta^2\rho_{k0}^2}{2s^4}\right)\right], \ \
\end{eqnarray}
where $\rho^2=x^2+y^2$ and $\rho_{k0}^2 = x^2+(y-id^2k_{0y})^2$.
It is seen that the packet, which was gaussian at $t=0$ (see Eq. (\ref{BG_psi})),
is {\it not} gaussian at later times (see Discussion). The upper and lower
components have the same decay time, oscillation period, etc. In order to characterize the
spreading (or dispersion) of the packet we calculate its width $\Delta R(t)$ as a function of
time
\begin{equation} \label{BGDeltaR2}
[\Delta R(t)]^2 = \langle \psi(\bm r,t)|\hat{\bm r}^2 - \langle \hat{\bm r} \rangle^2 | \psi(\bm r,t)\rangle,
\end{equation}
where $\psi(\bm r,t)$ is the above two-component wave function and
$\langle \hat{\bm r}\rangle = \langle\psi(\bm r,t)|\hat{\bm r}|\psi(\bm r,t) \rangle$.
The calculated width $\Delta R$ is plotted versus time in Fig. \ref{Fig1}d. It is seen
that during the initial 80 femtoseconds the packet's width increases
only twice compared to its initial value,
while the ZB and the accompanying effects disappear almost completely during this time. We conclude
that the spreading of the packet is {\it not} the main cause of the transient character of the ZB.
In fact, also the spreading oscillates a little, but this effect is too small to be
seen in Fig. \ref{Fig1}d.

It is well known that the phenomenon of ZB is due to an interference of wave functions
corresponding to positive and negative eigen-energies of the initial Hamiltonian.
Looking for physical reasons behind the transient character of ZB described above,
we decompose the total wave function $\psi(\bm r,t)$ into the positive ($p$) and
negative ($n$) components  $\psi^p(\bm r,t)$ and $\psi^n(\bm r,t)$.
We have
\begin{eqnarray} \label{BG_psit_S}
|\psi(t)\rangle &=& e^{-i\hat{H}t/\hbar} |\psi(0)\rangle  \nonumber \\
 &=& e^{-iEt/\hbar}\langle p|\psi(0)\rangle  |p\rangle +
      e^{iEt/\hbar}\langle n|\psi(0)\rangle  |n\rangle, \ \
\end{eqnarray}
where $|p\rangle$ and $|n\rangle$ are the eigen-functions of the Hamiltonian (\ref{BG_H})
in $\bm k$ space corresponding to positive and negative energies, respectively.
Further
\begin{eqnarray}
 \langle \bm k|p\rangle &=& \frac{1}{\sqrt{2}}\left(\begin{array}{c} 1 \\
            k_+^2/k^2  \end{array}\right) \delta(\bm k-\bm k^{'}),\\
 \langle \bm k|n\rangle &=& \frac{1}{\sqrt{2}}\left(\begin{array}{c} 1 \\
           -k_+^2/k^2  \end{array}\right)\delta(\bm k-\bm k^{'}).
\end{eqnarray}
After some manipulations we finally obtain
\begin{eqnarray} \label{BG_psit_pos}
 \psi^p(\bm r,t)= \frac{1}{4\pi}\frac{d}{\sqrt{\pi}}
    \int d^2 \bm k e^{-\frac{1}{2}d^2(k_x^2+(k_y-k_{0y})^2)} e^{i\bm k\bm r} e^{-iEt/\hbar}
    \times \nonumber \\ \ \ \
    \left(\begin{array}{c} 1 \\  k_+^2/k^2 \end{array}\right). \ \ \ \ \ \
\end{eqnarray}
The function $\psi^n(\bm r,t)$ is given by the identical expression with the changed
signs in front of $E$ and $k_+^2/k^2$ terms. There is
$\psi(\bm r,t)=\psi^p(\bm r,t)+\psi^n(\bm r,t)$ and $\langle \psi^n|\psi^p\rangle=0$.

Now we calculate the average values of $\bar{x}$ and $\bar{y}$ using the positive
and negative components in the above sense. We have
\begin{equation}
 \bar{x}(t) = \int (\psi^n + \psi^p)^{\dagger} x (\psi^n + \psi^p) d^2 \bm r,
\end{equation}
so that we deal with four integrals. A direct calculation gives
\begin{equation} \label{BG_x_mix}
 \int |\psi^p|^2x d^2 \bm r + \int |\psi^n|^2x d^2 \bm r = \bar{x}_c,
 \end{equation}
\begin{equation}
  \int \psi^{n\dagger}x \psi^p d^2 \bm r + \int \psi^{p\dagger}x \psi^n d^2 \bm r = \bar{x}_Z(t),
 \end{equation}
where $\bar{x}_c$ and $\bar{x}_Z(t)$ have been defined in Eq. (\ref{BG_x(t)}).
Thus the integrals involving only the positive and only the negative components
give the constant shift due to ZB, while the mixed terms lead to the ZB oscillations.
All terms together reconstruct the result (\ref{BG_x(t)}).

Next we calculate the average value $\bar{y}$. There is no symmetry between
$\bar{x}$ and  $\bar{y}$ because the wave packet is centered around
$k_x=0$ and $k_y=k_{0y}$. The average value  $\bar{y}$ is again given
by four integrals. However, now the mixed terms vanish since they
contain odd integrands of $k_x$, while the integrals involving the positive and negative
components alone give
\begin{eqnarray}
 \int |\psi^p|^2y d^2 \bm r &=& \frac{\hbar k_{0yt}}{2m^*}, \\
 \int |\psi^n|^2y d^2 \bm r &=& -\frac{\hbar k_{0yt}}{2m^*}.
\end{eqnarray}
This means that the 'positive' and 'negative' subpackets move in the opposite
directions with the same velocity $v=\hbar k_{0yt}/2m^*$. The relative velocity
is $v^{rel}=\hbar k_{0yt}/m^*$. Each of these packets has the initial width $d$
and it (slowly) spreads in time. After the time $\Gamma_Z^{-1}=d/v^{rel}$
the distance between the two packets equals $d$, so the integrals (\ref{BG_x_mix})
are small, resulting in the diminishing Zitterbewegung amplitude. This reasoning
gives the decay constant $\Gamma_Z=\hbar k_{0yt}/m^*d$, which is exactly
what we determined above from the analytical results (see Eq. (\ref{BG_GammaZ})).
Thus, {\it the transient character of the ZB oscillations is due to the
increasing spacial separation of the subpackets} corresponding to the positive and negative
energy states. This confirms our previous conclusion that it is not the packet's
slow spreading that is responsible for the attenuation  (see Discussion).
However, as we show below, also spreading may possibly play this role in some cases.

To conclude our analytical discussion of the ZB in bilayer graphene we consider
an interesting property of the velocity squared.
If $\hat{v}_x = \partial \hat{H}/\partial \hat{p}_x$
and $\hat{v}_y = \partial \hat{H}/\partial \hat{p}_y$ are calculated directly from
the Hamiltonian (\ref{BG_H}), then it is easy to show that $\hat{v}_x^2=\hbar^2k^2/m^2$ and
$\hat{v}_y^2=\hbar^2k^2/m^2$, so that $\hat{v}^2=\hat{v}_x^2+\hat{v}_y^2=2\hbar^2k^2/m^2$
does not depend on time. In the Heisenberg picture we split the velocity components
into 'classical' and ZB parts
\begin{eqnarray} \label{BG_vx(t)}
 \hat{v}_x^Z(t) &=&\frac{\hbar k_y}{mk^2} \left(\begin{array}{cc}
   \!\! k^2\sin(\hbar k^2t/m) & \!\!\ -ik_+^2\cos(\hbar k^2t/m) \\
  \!\! ik_-^2\cos(\hbar k^2t/m) & \!\! -k^2\sin(\hbar k^2t/m) \\
 \end{array}\right),
 \nonumber \\
 \hat{v}_x^c(t) &=&  \frac{\hbar k_x}{mk^2}
  \left(\begin{array}{cc}  0 & k_+^2 \\ k_-^2 &0 \\  \end{array}\right),
\end{eqnarray}
and similarly for $\hat{v}_y(t)$. Noting that $\{\hat{v}_x^Z(t),\hat{v}_x^c(t) \}=0$
we have $\hat{v}_x(t)^2=\hat{v}_x^Z(t)^2+\hat{v}_x^c(t)^2$. Using Eq. (\ref{BG_vx(t)})
we show that each of these terms is time independent:
$\hat{v}_x^Z(t)^2=\hbar^2k_y^2/m^2$ and $\hat{v}_x^c(t)^2=\hbar^2k_x^2/m^2$, and
similarly for $\hat{v}_y^Z(t)^2$ and $\hat{v}_y^c(t)^2$.
Thus, the velocity squared of the ZB component
$\hat{v}^Z(t)^2$=$\hat{v}_x^Z(t)^2+\hat{v}_y^Z(t)^2$=$\hbar^2k^2/m^2$
is equal to that of the 'classical' component
$\hat{v}^c(t)^2$=$\hat{v}_x^c(t)^2+\hat{v}_y^c(t)^2$=$\hbar^2k^2/m^2$.

\section{Monolayer graphene}

\begin{figure}
\includegraphics[width=8.5cm,height=8.5cm]{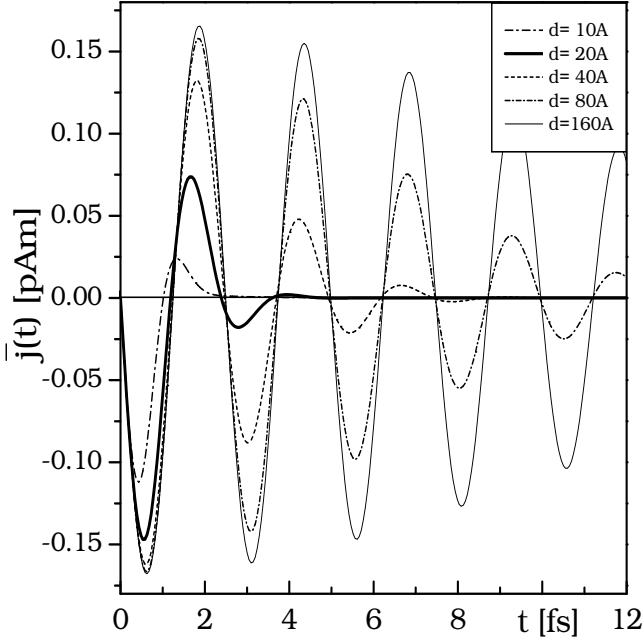}
\caption{ \label{Fig2} Oscillatory electric current caused by the ZB in monolayer graphene
{\it versus} time, calculated for a gaussian wave packet with $k_{0y}=1.2\times 10^9$m$^{-1}$ and various
packet widths $d$.}
\end{figure}

Now we turn to monolayer graphene. The two-dimensional band Hamiltonian
describing its band structure is
\cite{Wallace47,Slonczewski58,McClure56,Novoselov05,Zhang05,Sadowski06}
\begin{equation} \label{MG_H}
 \hat{H}_M = u\left(\begin{array}{cc}
     0 & p_x-ip_y \\  p_x+ip_y & 0 \\     \end{array}\right),
\end{equation}
where $u\approx 1\times 10^8$cm/s. The resulting energy dispersion is linear in momentum:
${\cal E}=\pm u\hbar k$, where $k=\sqrt{k_x^2+k_y^2}$.
The quantum velocity in the Schroedinger picture is
$\hat{v}_i=\partial H_M/\partial \hat{p}_i$, it does not commute with the Hamiltonian (\ref{MG_H}).
In the Heisenberg picture we have
$\hat{\bm v}(t)=\exp(i\hat{H}_Mt/\hbar)\hat{\bm v}\exp(-i\hat{H}_Mt/\hbar)$.
Using Eq. (\ref{MG_H}) we calculate
\begin{equation} \label{MG_v11}
 \hat{v}_x^{(11)} = u \frac{k_y}{k}\sin(2ukt).
\end{equation}
The above equation describes the trembling motion with the frequency $\omega_Z=2uk$,
determined by the energy difference between the upper and lower energy branches for a given
value of $k$. As before, the ZB in the direction $x$ occurs only if there is a non-vanishing
momentum $\hbar k_y$. We calculate an average velocity (or current) taken over the wave
packet given by Eq. (\ref{BG_psi}). The averaging procedure amounts to a double integral.
The latter is not analytical and we compute it numerically. The results for the current
$\bar{j}_x=e\bar{v}_x$ are plotted in Fig. \ref{Fig2} for $k_{0y}=1.2\times 10^9$m$^{-1}$
and different realistic packet widths $d$ (see Ref. \cite{Schliemann07}).
It is seen that the ZB frequency does not depend
on $d$ and is nearly equal to $\omega_Z$ given above for the plane wave.
On the other hand, the amplitude of ZB does depend on $d$ and we deal with decay times
of the order of femtoseconds. For small $d$ there are almost no oscillations, for
very large $d$ the ZB oscillations are undamped. These conclusions agree with our analytical
results for bilayer graphene. The behavior of ZB depends quite critically on the values of
$k_{0y}$ and $d$, which is reminiscent of the damped harmonic oscillator.

\section{Carbon nanotubes}
\begin{figure}
\includegraphics[width=8.5cm,height=8.5cm]{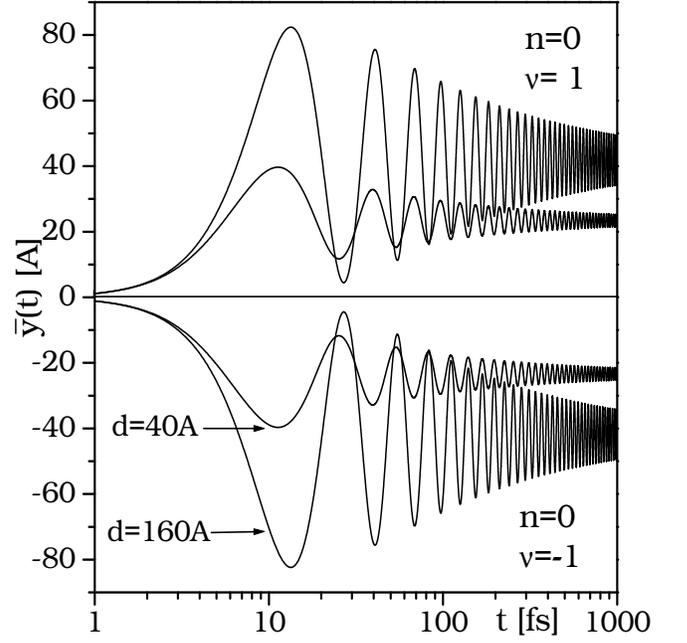}
\caption{ \label{Fig3} Zitterbewegung of two charge carriers in the ground subband of a single carbon
nanotube of $L=200$ \AA\ {\it versus} time (logarithmic scale),
calculated for gaussian wave packets of two different
widths $d$. After the ZB disappears a constant shift remains.
The two carriers are described by different quantum numbers $\nu$.
At higher times the amplitude of ZB oscillations decays as $t^{-1/2}$ (see text).}
\end{figure}

Finally, we consider monolayer graphene sheets rolled into single
semiconducting carbon nanotubes (CNT). The band Hamiltonian in
this case is similar to Eq. (\ref{MG_H}) except that, because of
the periodic boundary conditions, the momentum $p_x$ is quantized
and takes discrete values $\hbar
k_x=\hbar k_{n\nu}$, where $k_{n\nu}=(2\pi/L)(n-\nu/3)$, $n=0,\pm 1,\ldots$,
$\nu=\pm 1$, and $L$ is the length of circumference of CNT \cite{Saito99,Ando93}.
As a result, the free electron motion can occur only in the direction
$y$, parallel to the tube axis. The geometry of CNT has two
important consequences. First, for $\nu=\pm 1$ there {\it always}
exists a non-vanishing value of the quantized momentum $\hbar k_{n\nu}$.
Second, for each value of $k_{n\nu}$ there
exists $k_{-n,-\nu}=-k_{n\nu}$ resulting in the
same subband energy ${\cal E}=\pm E$, where
\begin{equation} \label{CNT_E}
 E=\hbar u\sqrt{k_{n\nu}^2+k_y^2}.
\end{equation}
The time dependent velocity $\hat{v}_y(t)$ and the displacement $\hat{y}(t)$ can be
calculated for the plane electron wave in the usual way and they exhibit the ZB
oscillations (see Ref. \cite{Zawadzki06}). For small momenta $k_y$ the ZB frequency
is $\hbar\omega_Z=E_g$, where $E_g=2\hbar uk_{n\nu}$.
The ZB amplitude is $\lambda_Z\approx 1/k_{n\nu}$.
However, we are again interested in the displacement $\bar{y}(t)$ of a charge carrier
represented by a one-dimensional wave packet analogous to that described in Eq. (\ref{BG_psi})
\begin{equation} \label{CNT_psi}
 \psi(y)= \frac{1}{\sqrt{2\pi}}\frac{d^{1/2}}{\pi^{1/4}} \int dk_y e^{-\frac{1}{2}d^2k_y^2} e^{ik_yy}
 \left(\begin{array}{c} 1 \\ 0 \end{array}\right).
\end{equation}
The average displacement is $\bar{y}(t)=\bar{y}_Z(t)-\bar{y}_{sh}$ where
\begin{equation} \label{CNT_y_osc}
\bar{y}_Z(t) = \frac{\hbar^2du^2k_{n\nu}}{2\sqrt{\pi}} \int_{-\infty}^{\infty}
    \frac{dk_y} {E^2}\cos\left(\frac{2Et}{\hbar}\right)e^{-d^2k_y^2}
\end{equation}
and $\bar{y}_{sh}= 1/2\sqrt{\pi}d\ {\rm sgn}(b)[1-\Phi(|b|)]\exp(b^2)$, where $b=k_{n\nu}d$ and
$\Phi(x)$ is the error function. The ZB oscillations of $\bar{y}(t)$ are plotted in Fig. \ref{Fig3}
for $n=0$, $\nu=\pm 1$ and $L=200$ \AA.
It can be seen that, after the transient ZB oscillations disappear, there
remains a shift $\bar{y}_{sh}$. Thus the ZB separates spatially the charge carriers that are
degenerate in energy but characterized by $n,\nu$ and $-n,-\nu$ quantum numbers.
The current is proportional to $\bar{v}_y=\partial \bar{y}/\partial t$, so that the currents related to
$\nu=1$ and $\nu=-1$ cancel each other. To have a non-vanishing current one needs to break the
above symmetry, which can be achieved by applying an external magnetic field
parallel to the tube axis \cite{Zawadzki06}.

Above we considered the situation in which a non-vanishing  value of transverse momentum
$\hbar k_x$ is 'built in' by the tube topology. However, it is also possible to prepare a
wave packet with an initial non-vanishing momentum $k_{0y}$. Using the method presented above
for bilayer graphene (see Eq. (\ref{BG_psit_S})) we can decompose the total wave packet (\ref{CNT_psi})
into the positive and negative subpackets with the result
\begin{equation} \label{CNT_psit_pos}
 \psi^p(y,t)= \frac{d^{1/2}}{2^{3/2}\pi^{3/4}}
   \! \int \! dk_y e^{-\frac{1}{2}d^2(k_y-k_{0y})^2}\! e^{ik_yy} e^{-iEt}\!
 \left(\begin{array}{c} 1 \\ \!\!k_+/k\!\! \end{array}\right).
\end{equation}
The function $\psi^n(y,t)$ is given by a similar expression with the
changed signs in front of $E$  and $(k_+/k)$ terms.
Here we use the notation $k=\sqrt{k_{n\nu}^2+k_y^2}$ and $k_+=k_{n\nu}+ik_y$.
Now the oscillating part of $\bar{y}$ is, as before
\begin{equation}
  \int \psi^{n\dagger}y \psi^p dy + \int \psi^{p\dagger}y \psi^n dy = \bar{y}_Z(t,k_{0y}).
\end{equation}
For $k_{0y}=0$ the above $\bar{y}_Z(t,k_{0y})$ reduces to  $\bar{y}_Z(t)$
given by Eq. (\ref{CNT_y_osc}). The average contributions of positive (or negative)
terms alone are
\begin{equation} \label{CNT_psip2y}
  \int |\psi^p_{(n)}|^2ydy = \frac{1}{2}\bar{y}_c \pm ut \int \frac{k_y|F(k_y)|^2dk_y}{2\sqrt{k_{n\nu}^2+k_y^2}},
\end{equation}
where $F(k_y) =d^{1/2}/(2\pi^{1/4})
\exp(-\frac{1}{2}d^2(k_y-k_{0y})^2)$ is the packet function. The sum
of the first terms for $\psi^p(y,t)$ and $\psi^n(y,t)$ in Eq.
(\ref{CNT_psip2y}) gives $\bar{y}_c$, as before. For $k_{0y}=0$ the
second term vanishes which physically means that the relative
velocity of the two subpackets is zero, so that they stay together
in time. It is for this reason that the decay of ZB is slow (see
Fig. \ref{Fig3}). If $k_{0y}\neq 0$, the second term in Eq.
(\ref{CNT_psip2y}) does not vanish, the two subpackets run away from
each other, their overlap diminishes and the ZB disappears much more
quickly.

The question remains: what is the physical reason for the (slow) damping of the ZB electron
shown in Fig. {\ref{Fig3}}, if the subpackets stay together? (As we mentioned in the Introduction,
the mathematical expression for the damping phenomenon is the Riemann-Lebesgue theorem.)
Trying to answer this question we calculated the spreading of the wave subpackets (\ref{CNT_psit_pos})
in time. For the initial width $\Delta y \approx 90$ \AA\ the subpackets reach
the width $\Delta y \approx 2600$ \AA\ after the time of $1000$ fs (see Fig. \ref{Fig3}). Thus,
we would be tempted to say that it is the spreading of the packets that is responsible for the
attenuation of ZB. However, it should be noted that, while at higher times the packet's
dispersion is linear in time (see Ref. \cite{Garraway95} and Fig. \ref{Fig3}d),
the amplitude of ZB oscillations
decays as $t^{-1/2}$. A similar slow damping of ZB occurs for
one-dimensional relativistic electrons in a vacuum
if the average momentum of the subpackets is zero (see Discussion).

\section{Discussion}
It is of interest that the ZB phenomena similar to those described above occur
also for wave packets representing free relativistic electrons in a vacuum
governed by the Dirac equation. This confirms again the strong similarity of the
two-band models for non-relativistic electrons
in solids to the description of free relativistic electrons in a vacuum,
see Refs. \cite{Zawadzki05KP,Zawadzki06,Rusin07,ZawadzkiOPS,ZawadzkiHMF}.
In contrast to bilayer graphene, the kinematics of the one-dimensional
relativistic wave packets may not be described analytically, so the
solutions were computed numerically and visualized graphically by Thaller \cite{Thaller04}.
It was shown that: 1) An initial relativistic gaussian wave packet after spreading
is not gaussian any more. This is analogous to our Eqs. (\ref{BG_psi_up}) and (\ref{BG_psi_low}).
2) If an average momentum of the initial positive and negative subpackets is zero,
the overlap of the two subpackets  remains almost constant in time
and the resulting ZB decays quite slowly.
This corresponds to our considerations of CNT with $k_{0y}=0$, see Fig. \ref{Fig3}.
(It is to be reminded that the two overlapping subpackets are orthogonal to each other.)
3) If the initial average momentum of both subpackets
is nonzero, the two subpackets quickly run away from each other and the ZB falls quickly
since it is sustained only when the subpackets have some overlap in the position
space. This corresponds to our considerations of bilayer graphene, see Fig. \ref{Fig1}.

The transient ZB of free relativistic wave packets in a vacuum was also studied numerically
by Braun {\it et al.} \cite{Braun99}. It was shown that, for example, the decay times
of a typical wave packet having the width $\Delta x=\lambda_c$ and the
initial wave vector $k_{0x}=1.37$ a.u. is $2.4\times 10^{-5}$ fs.
This should be compared with our predicted decay times of $\Gamma_Z^{-1}=40$ fs for bilayer graphene.
It turns out once again that solids are much more promising media for an observation of
Zitterbewegung than a vacuum.

The Zitterbewegung phenomenon described above should not be confused with the
Bloch oscillations of charge carriers in superlattices, although the latter occur at
picosecond frequencies and have comparable picoseconds decay times
(see e.g. Refs. \cite{Martini96,Lyssenko97,Kosevich06}).
However, the Bloch oscillations are basically a one-band phenomenon, they have been realized
in superlattices (although this is in principle not the condition {\it sine qua non})
and, most importantly, they require an external electric field driving electrons all
the way to the Brillouin zone boundary. On the other hand, the ZB needs at least two bands
and it is a no-field phenomenon. On the other hand, narrow-gap superlattices could
provide a suitable medium of its observation.

In view of our results it is clear that, in order to observe the transient Zitterbewegung, it is
necessary to prepare simultaneously a sufficient number of charge carriers in the form of wave
packets. If one wants to detect the current, the trembling motion of all carriers must have the same
phase. On the other hand, if one wants to see only the remnant displacement, the phase coherence is
not necessary. As we said above, the ZB frequency is to a good accuracy given by the corresponding
energy difference between the upper and lower energy branches
while the amplitude depends strongly on packet's
width. For the two graphene materials considered above one needs an initial momentum in one
direction to have the ZB along the transverse direction (see also Ref. \cite{Schliemann05}).
For nanotubes the initial momentum is automatically there due to the circular boundary conditions.
The oscillatory motion between two graphene layers, as illustrated in Fig. \ref{Fig1}c, appears
to be a promising phenomenon for observation. As far as the detection is concerned, one needs sensitive
current meters or scanning probe microscopy, both working at infrared frequencies
and femtosecond to picosecond decay times (see Refs. \cite{Topinka00,LeRoy03}).

\section{Summary}
In summary, using the two-band structure of bilayer graphene, monolayer graphene and carbon nanotubes
we show that charge carriers in these materials, localized in the form of gaussian wave packets,
exhibit the transient Zitterbewegung with the decay times of femtoseconds in graphene and
picoseconds in nanotubes. Observable dynamical ZB effects, most notably the electric current, are
described. It is demonstrated that, after the trembling motion disappears, there remains its 'trace' in
the form of a persistent charge displacement.
It is emphasized that the described ZB in solids is in close analogy to that of the relativistic
electron in a vacuum.

\acknowledgements
We acknowledge elucidating discussions with Professor I. Birula-Bialynicki.
This work was supported in part by the Polish Ministry of Sciences,
Grant No PBZ-MIN-008/P03/2003.

\end{document}